\documentclass[twoside,slac_one]{revtex4}
\usepackage{graphicx}
\usepackage{fancyhdr}
\usepackage{amsmath} 
\usepackage{bm}
\usepackage{amsxtra}
\usepackage{amssymb}
\usepackage{amsthm}
\usepackage{latexsym}
\usepackage{lscape}

\usepackage{multirow}
\usepackage{hhline}
\usepackage{epstopdf}

\begin{document}

\title{Search for Electron Neutrino Appearance in MINOS}

%

\author{M. Orchanian, on behalf of the MINOS Collaboration}
\affiliation{Division of Physics, Mathematics, and Astronomy, California Institute of Technology, Pasadena, CA, USA}

\begin{abstract}
The MINOS Collaboration continues its search for $\nu_e$ appearance in the NuMI (Neutrinos at the Main Injector) beam at Fermilab. Neutrinos in the beam interact in the Near Detector, located 1 km from the beam source, allowing us to characterize the backgrounds present in our analysis. In particular, we can estimate the number of $\nu_e$ candidate events we expect to see in the Far Detector (735 km away, in the Soudan mine in northern Minnesota) in the presence or absence of $\nu_\mu \rightarrow \nu_e$ oscillation. Recent efforts to improve the sensitivity of the analysis, including upgrades to the event identification algorithm and fitting procedure, are discussed, and the latest results from the search are presented.
\end{abstract}

\maketitle

\thispagestyle{fancy}


\section{Introduction and Motivation}
\begin{figure}[ht]
\centering
\includegraphics[width=80mm]{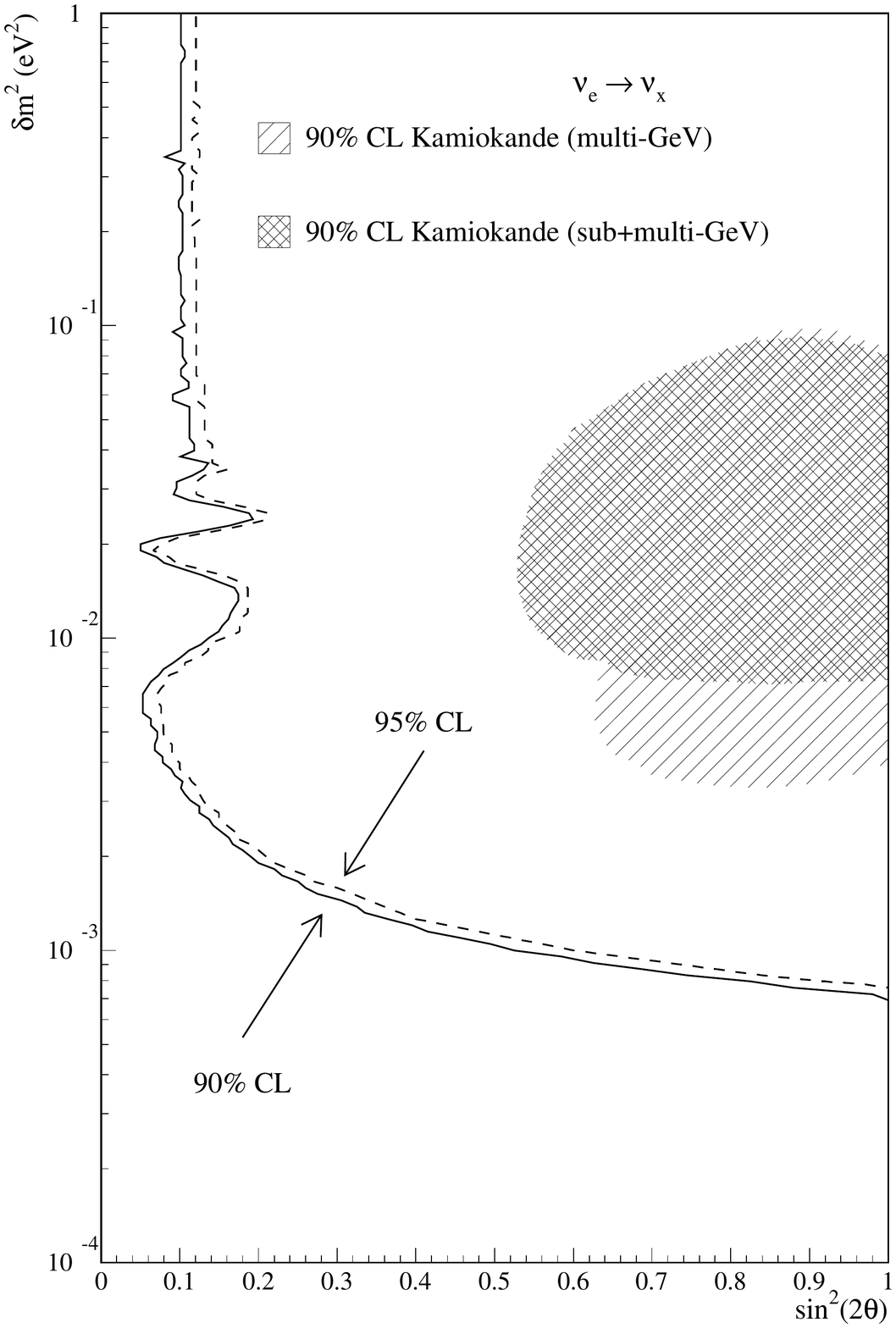}
\caption{Exclusion contours from the CHOOZ experiment constraining the value of the $\theta_{13}$ mixing angle (horizontal axis) as a function of $\Delta m_{32}^2$ (vertical axis). Taken from Ref.~\cite{CHOOZpaper}.} \label{CHOOZplot}
\end{figure}

Neutrino oscillation is now an established phenomenon, thanks to a variety of experiments~\cite{pastexpt1,pastexpt2,pastexpt3,pastexpt4,pastexpt5,pastexpt6,pastexpt7} which have probed various apparent anomalies in expected neutrino spectra from the sun, atmosphere, nuclear reactors, and accelerator-based beams. The three weak eigenstates -- $\nu_e$, $\nu_\mu$, $\nu_\tau$ -- are related to the three mass eigenstates -- $\nu_1$, $\nu_2$, $\nu_3$ -- via the PMNS (for Pontecorvo, Maki, Nakagawa, and Sakata) matrix, a unitary mixing matrix $U$ parametrized by three mixing angles -- $\theta_{12}$, $\theta_{13}$, $\theta_{23}$ -- and three CP-violating phases -- $\delta$, $\alpha_1$, $\alpha_2$~\cite{pontecorvo}. No hints exist yet as to the value of $\delta$, and oscillation experiments are not sensitive to the latter two CP-violating phases, which are nonzero only if neutrinos are their own antiparticles (i.e., Majorana particles). So far, two of the mixing angles, $\theta_{12}$ and $\theta_{23}$, have been measured~\cite{pastexpt1,pastexpt4,pastexpt6} and are large. The value of $\theta_{13}$, on the other hand, appears to be quite small. It was first significantly constrained by the CHOOZ experiment in 1999~\cite{CHOOZpaper}; the result is shown in Fig. ~\ref{CHOOZplot}. These proceedings report the latest improvements on this constraint by the MINOS experiment.

The probability of a neutrino of energy $E$ created with flavor $\alpha$ being detected with flavor $\beta$ after traveling some distance $L$ in vacuum -- to wit, the oscillation probability -- is given by:

\begin{equation}
P(\nu_\alpha \rightarrow \nu_\beta) = \left|\sum\limits_{j=1}^3 U^{*}_{\alpha j}\exp\left(-\frac{im_{j}^{2}L}{2E}\right)U_{\beta j}\right|^2,
\label{oscprobgen}
\end{equation}

\noindent where $m_j$ is the mass of the $j$-th neutrino mass eigenstate. The final expression is found to depend not on the absolute squared masses but on two independent squared-mass differences -- $\Delta m_{21}^2$ and $\Delta m_{32}^2$. While both the sign and magnitude of the first of these are known~\cite{pastexpt1}, only the magnitude of the second squared-mass difference is known~\cite{pastexpt6}. That is, we do not know if $\nu_3$ is more or less massive than $\nu_1$ and $\nu_2$; the first possibility is referred to as the normal mass hierarchy and the latter, the inverted mass hierarchy.

The oscillation mode of present interest is $\nu_\mu \rightarrow \nu_e$. To leading order:

\begin{equation}
P(\nu_\mu \rightarrow \nu_e) \approx \sin^2(2\theta_{13})\sin^2(\theta_{23})\sin^2(1.27\Delta m^2_{32}L/E),
\label{oscprobleading}
\end{equation}

\noindent with $\Delta m^2_{32}$ in units of $\mathrm{eV}^2$, $L$ in km, and $E$ in GeV. When we account for the propagation of the neutrinos through matter, additional, higher order terms appear in Eq. ~\ref{oscprobleading}~\cite{akhmedov}. The full expression depends on the CP-violating phase $\delta$ and the mass hierarchy; accordingly, the results presented here will be sensitive to these parameters.

Under certain conditions~\cite{sakharov}, leptonic CP violation can lead to non-conservation of baryon number, which in turn can help explain the observed matter-antimatter asymmetry in our universe. Access to the value of the leptonic CP-violating phase, namely $\delta$, is controlled by the size of $\theta_{13}$. Should $\theta_{13}$ be sufficiently large, one may be able to measure $\delta$ with current or next-generation long-baseline neutrino oscillation experiments. Finding, on the other hand, that $\theta_{13}$ is zero could point to a new symmetry.

\section{The MINOS Experiment}
The MINOS experiment consists of three key parts: the Neutrinos at the Main Injector (NuMI) neutrino beam at Fermilab, the Near Detector (ND) at Fermilab, and the Far Detector (FD) in Soudan, MN. The NuMI beam~\cite{numi} is produced by striking a graphite target with 120-GeV protons from Fermilab's Main Injector, focusing the resulting secondary hadrons (mostly pions, some kaons) using pulsed-current magnetic horns, and allowing these hadrons to decay leptonically in flight. Although the horns' polarities are set to focus $\pi^{+}$ and $K^{+}$ and defocus $\pi^{-}$ and $K^{-}$, a small fraction of negative hadrons travel along the horns' axis, avoid being defocused, and remain in the beam. The primary decay mode for these hadrons is $\mu^{+}\nu_\mu$ or $\mu^{-}\bar{\nu}_\mu$; a very small fraction decay to $e^{+}\nu_e$ or $e^{-}\bar{\nu}_e$. Ultimately, the resulting neutrino beam is 92\% $\nu_\mu$ and 7\% $\bar{\nu}_\mu$; $\nu_e$ and $\bar{\nu}_e$ make up the remainder.

The MINOS detectors are functionally identical, magnetized tracking calorimeters. The 980-ton ND is located approximately 1 km downstream of the NuMI target and measures the neutrino beam's initial composition and energy spectrum. The 5.4-kton FD is located 735 km downstream in the Soudan mine in northern Minnesota and measures the oscillated beam's composition and energy spectrum. Both detectors consist of alternating layers of 2.54-cm thick steel and 1-cm thick plastic scintillator. The scintillator is segmented into optically-isolated, 4.1-cm wide strips and serves as the active portion of each detector. Light from the strips is read out via optical fibers and multi-anode photomultiplier tubes. Further details can be found in Ref.~\cite{minosnim}.

\section{Neutrino Interactions in the MINOS Detectors}

\subsection{Event Types}

\begin{figure*}[ht]
\centering
\includegraphics[width=135mm]{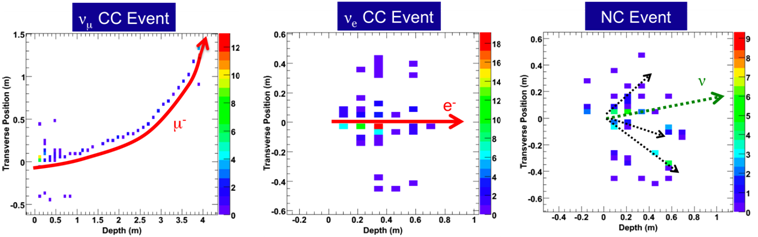}
\caption{Displays of simulated events in the MINOS detectors. Pulse heights shown are calibrated such that 1 unit is the amount of energy deposited by a minimum ionizing particle.} \label{eventdisplay}
\end{figure*}

Though a multi-purpose experiment, MINOS was designed and optimized to study $\nu_\mu$ disappearance. The event type of interest in such an analysis is a $\nu_\mu$ charged-current (CC) interaction:

\begin{equation}
\nu_\mu + N \rightarrow \mu^{-} + X,
\label{numuccrxn}
\end{equation}

\noindent where N represents a nucleon or a quark inside a nucleon and X represents the resulting hadronic recoil system. The track produced by the muon typically extends far beyond the shower activity of the recoil system (see left of Fig.~\ref{eventdisplay}), enabling straightforward identification of such an event. In the analysis at hand, however, the event type of interest is a $\nu_e$ CC interaction:

\begin{equation}
\nu_e + N \rightarrow e^{-} + X.
\label{nueccrxn}
\end{equation}

\noindent In the MINOS detectors, the electromagnetic shower generated by the electron has a radiation length of 1.76 cm -- $\sim$30\% less than the thickness of a steel layer -- and a Moli\`{e}re radius of 3.7 cm -- $\sim$10\% less than the width of a scintillator strip. As a result, the electromagnetic shower becomes fully interspersed with the hadronic shower (see center of Fig.~\ref{eventdisplay}). The challenge, then, is to distinguish this type of event from the primary background event type -- the neutral-current (NC) interaction (see right of Fig.~\ref{eventdisplay}):

\begin{equation}
\nu_x + N \rightarrow \nu_x + X.
\label{ncrxn}
\end{equation}

\subsection{Event Identification}

This is the third $\nu_e$ appearance search conducted by MINOS. In all three analyses, event selection consisted of three general stages. First, we required that the event took place within a fiducial volume. Next,  we rejected events that were clearly or likely background. This ``pre-selection'' eliminated most $\nu_\mu$ CC events by requiring that any reconstructed track in the event be no more than 24 planes long and extend no more than 15 planes beyond the hadronic shower. We also required that an event have at least five contiguous planes with energy greater than half that deposited on average by a minimum ionizing particle. Events were required to be in-time with and to point in a direction consistent with the neutrino beam pulses. The reconstructed energy of the event had to lie between 1 and 8 GeV, as most events below 1 GeV are from NC interactions and $P(\nu_\mu \rightarrow \nu_e)$ is negligibly small above 8 GeV.

\begin{figure}[ht]
\centering
\includegraphics[width=80mm]{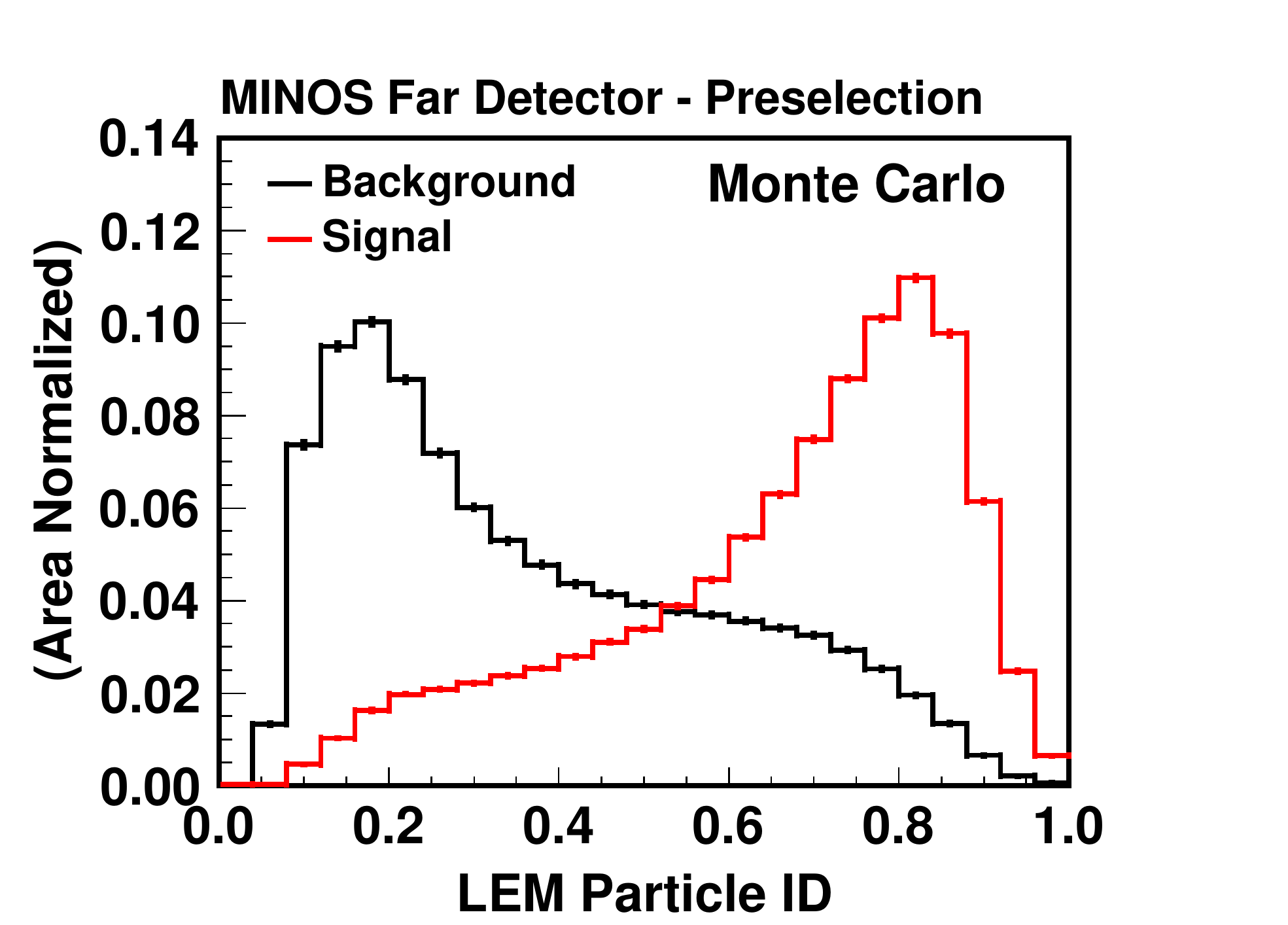}
\caption{LEM discriminant distributions for signal and background events passing the fiducial and pre-selection cuts.} \label{LEMdistro}
\end{figure}

The last stage of the selection involved the use of a sophisticated event identification technique to separate signal from background. In our first two $\nu_e$ appearance analyses~\cite{MINOSnue1,MINOSnue2}, we used a neural network (named Artificial Neural Network, or ANN) whose input variables were reconstructed quantities characterizing the longitudinal and transverse energy deposition profiles of an event. The present analysis utilizes instead a novel technique named Library Event Matching (LEM) that uses raw energy deposition information instead of reconstructed quantities and is based on a pattern recognition algorithm. We begin by comparing a given candidate event to each of $5\times 10^7$ simulated events (the ``library''), of which $\sim 3 \times 10^7$ are background events and $\sim 2 \times 10^7$ are signal events. Intuitively, the comparison determines the extent to which the topologies and strip pulse heights of the candidate and library events are similar. Quantitatively, this is performed using a likelihood:

\begin{equation}
  \log \mathcal{L} = \sum \limits_{j=1}^{N_{strips}} \sum \limits_{k=1}^{N_{planes}} \log \left[ \int^\infty_0 P(n_{cand}^{jk},\lambda) P(n_{lib}^{jk},\lambda) \, d \lambda \right],
\label{LEMlikelihood}
\end{equation}

\noindent where $P(n,\lambda)$ is the Poisson probability for observing $n$ given mean $\lambda$ and $n_x^{jk}$ is the charge (in photoelectrons) deposited in strip $j$ of plane $k$ of event $x$ (candidate or library). Once this matching process is completed for a given candidate event, the library events are ranked by their $\log \mathcal{L}$ values. The top 50 of these, designated the ``best matches,'' are summarized by three variables: (1) the fraction that are signal ($\nu_e$ CC) events, (2) for these signal events, the average fraction of total event energy in the hadronic recoil system, and (3) for these signal events, the average fraction of deposited charge that overlaps between each of them and the candidate event. These variables, along with the reconstructed energy of the candidate event, are given as inputs to a neural network; the output is the LEM discriminant, shown in Fig.~\ref{LEMdistro}.

\section{The Analysis}

\subsection{Overview}

\begin{figure}[ht]
\centering
\includegraphics[width=80mm]{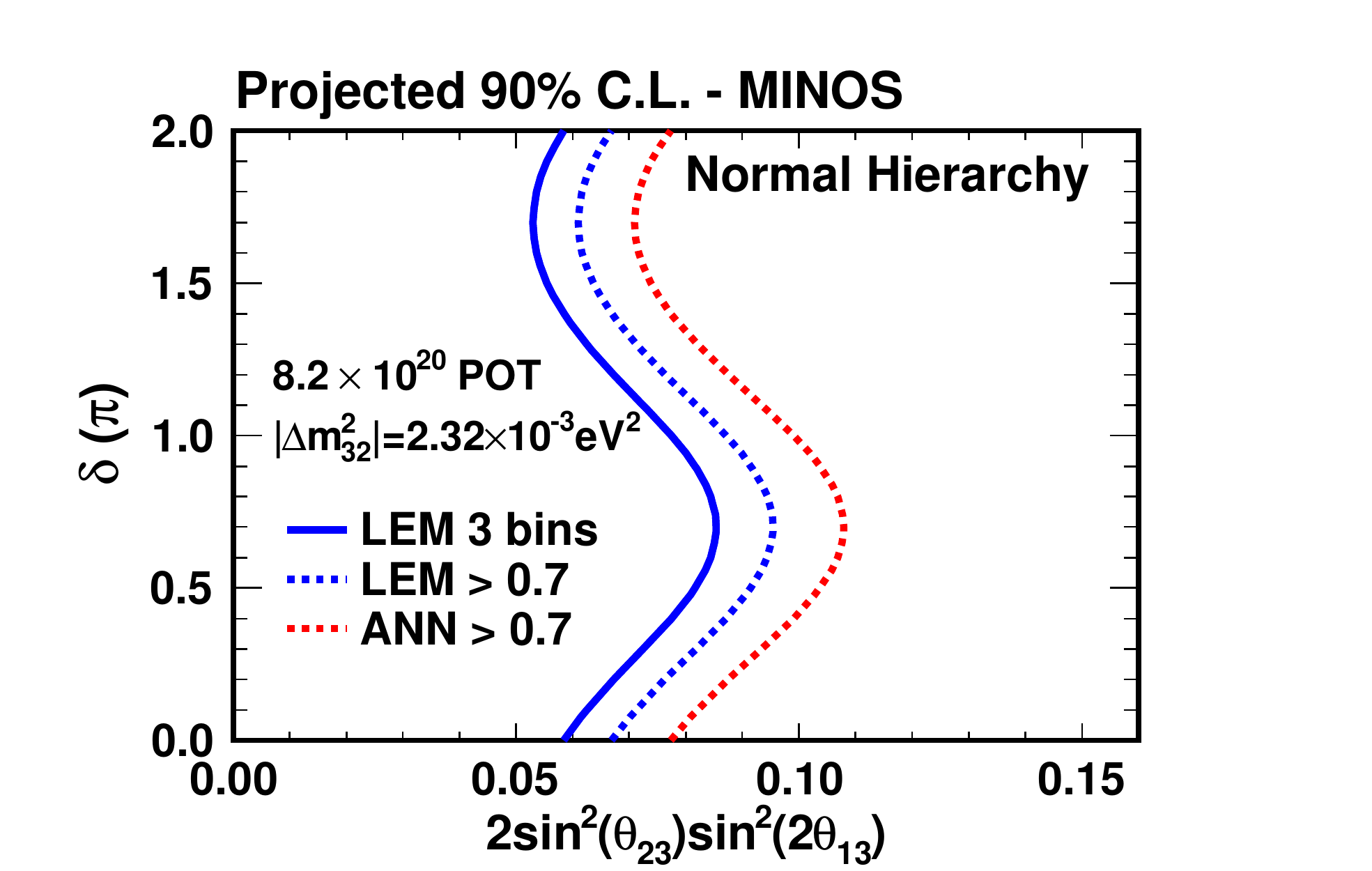}
\caption{Projected 90\% C.L. sensitivities to $\theta_{13} = 0$ assuming the normal mass hierarchy. The contour labeled ``ANN $>$ 0.7'' is the sensitivity achieved by repeating the last analysis with the additional data included. The ``LEM $>$ 0.7'' contour is the sensitivity for a counting experiment performed with LEM instead of ANN using the same full data set. Finally, the ``LEM 3 bins'' sensitivity is obtained by performing a shape fit to the LEM distribution.} \label{bettersens}
\end{figure}

The present analysis represents three improvements over the most recent previous analysis, the effects of which are summarized by Fig.~\ref{bettersens}. An additional exposure of $1.2 \times 10^{20}$ protons on the NuMI target provides a modest 17\% increase in the size of our data set. Using LEM instead of ANN provides a gain in sensitivity to $\theta_{13} = 0$ of about 15\%. Finally, instead of cutting at the optimal value of the discriminant (found to be 0.7 for LEM) and comparing the number of events above that value to the expectation (as we did in the previous two analyses), we perform a fit to the LEM discriminant distribution (optimal binning was found to be 0.6-0.7, 0.7-0.8, 0.8-1.0), obtaining an additional 12\% increase in sensitivity. The basic approach to making a background prediction, however, remains unchanged -- count the events observed in the ND and use the FD and ND simulations to ``extrapolate'' this to the FD.

\subsection{Background Prediction}

Events due to NC interactions constitute the primary background in this analysis. Contributions to the background from $\nu_\mu$ CC and intrinsic beam $\nu_e$ CC interactions are smaller but non-negligible; the $\nu_\mu \rightarrow \nu_\tau$ oscillation mode allows the possibility of $\nu_\tau$ CC interactions, an additional source of background. Such a multi-component background complicates the aforementioned extrapolation process, as each of the components is affected differently by the transition from ND to FD (e.g., NC interaction rates are not affected by oscillations while the various CC interaction rates are affected to differing extents). Consequently, we perform the background prediction component-by-component. In particular, the background prediction for component $\alpha$ (NC, $\nu_\mu$ CC, $\nu_e$ CC) in analysis bin $i$ is given by:

\begin{equation}
FD_{\alpha,i}^{Predicted} = ND_{\alpha,i}^{Data} \times \frac{FD_{\alpha,i}^{Simulation}}{ND_{\alpha,i}^{Simulation}},
\label{extrapolation}
\end{equation}

\noindent where the individual $FD$ and $ND$ terms stand for event counts. In a manner similar to that of the $\nu_\tau$ CC interactions, signal $\nu_e$ CC interactions ``appear'' due to oscillations (in this case, via the $\nu_\mu \rightarrow \nu_e$ mode). As such, these event types appear only in the FD and cannot be extrapolated from the ND using Eq.~\ref{extrapolation}; we use a more involved but conceptually similar approach based on the observed ND $\nu_\mu$ CC spectrum to predict the event counts due to these interactions.

\begin{figure}[ht]
\centering
\includegraphics[width=80mm]{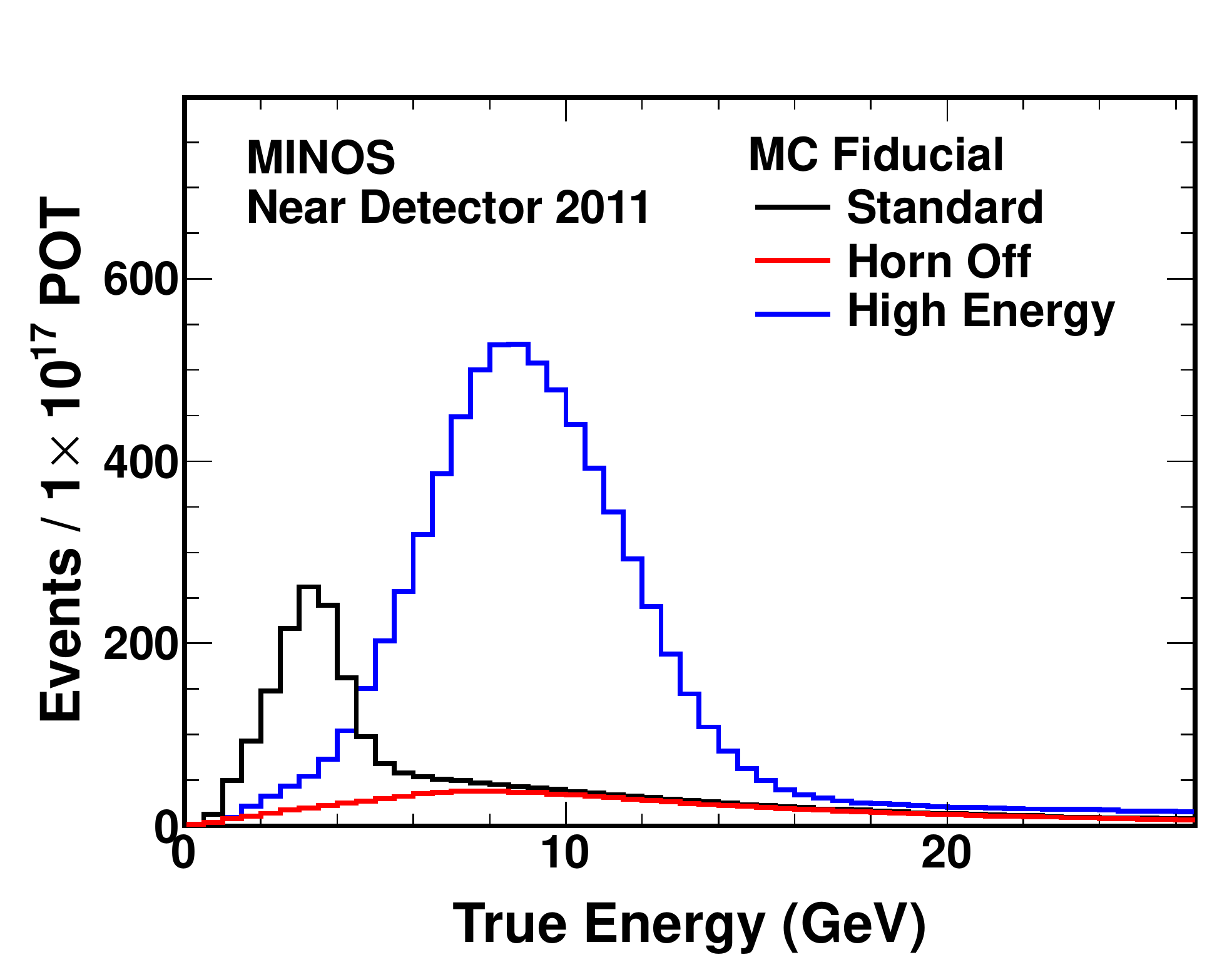}
\caption{Exposure-normalized energy spectra of all events occurring in the ND fiducial volume, for each of three different beam configurations described above.} \label{hornconfs}
\end{figure}

The $ND_{\alpha,i}^{Data}$ term stands out on the right-hand-side of Eq.~\ref{extrapolation}. While the total number of events in the ND data is straightforwardly obtained (simply apply the relevant cuts and count them), the means of determining the individual contributions of the three background event types to this number are far from obvious. Our solution to this problem takes advantage of the flexibility of the NuMI beam configuration. In addition to taking data in the standard configuration, we also take data (albeit with considerably smaller exposures) in the so-called ``horn-off'' (focusing horns turned off) and ``high-energy'' (target pulled farther away from the focusing horns, horn currents increased slightly) configurations. These configurations produce neutrino beams with different true energy spectra (see Fig.~\ref{hornconfs}), which in turn produce activity in the ND with different relative fractions of NC, $\nu_\mu$ CC, and intrinsic beam $\nu_e$ CC events.

\begin{figure}[ht]
\centering
\includegraphics[width=80mm]{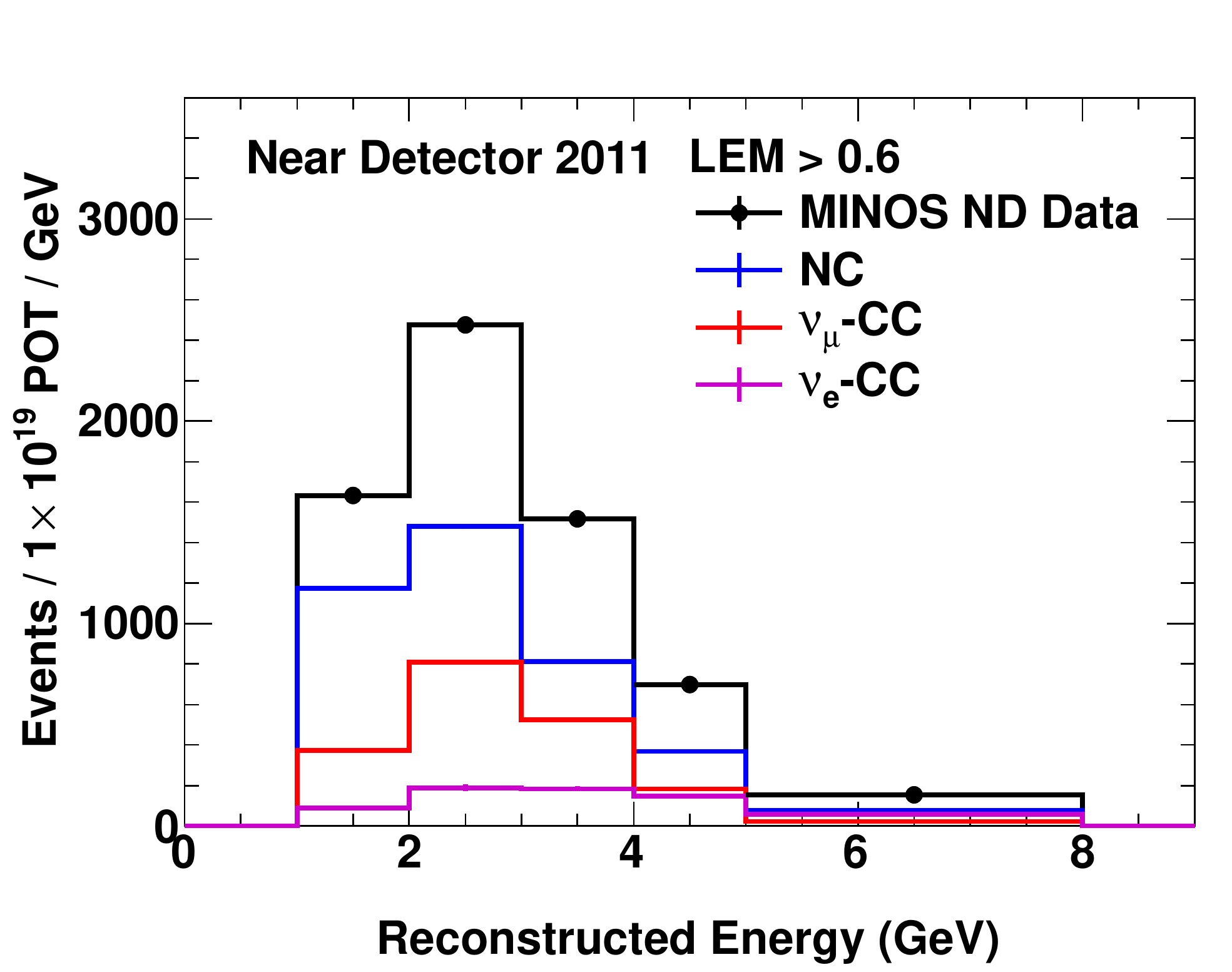}
\caption{The energy spectrum of all events in the ND used in the official analysis (i.e., those with LEM $>$ 0.6 and passing the fiducial and pre-selection cuts). The NC, $\nu_\mu$ CC, and $\nu_e$ CC spectra are obtained by the technique described above.} \label{nddecomp}
\end{figure}

Combining the total ND data rate in each of the three configurations with the simulated background component ratios between configurations, we perform a fit and obtain the $ND_{\alpha,i}^{Data}$, the data background components in the standard configuration, as shown in Fig.~\ref{nddecomp}.

\begin{figure}[ht]
\centering
\includegraphics[width=80mm]{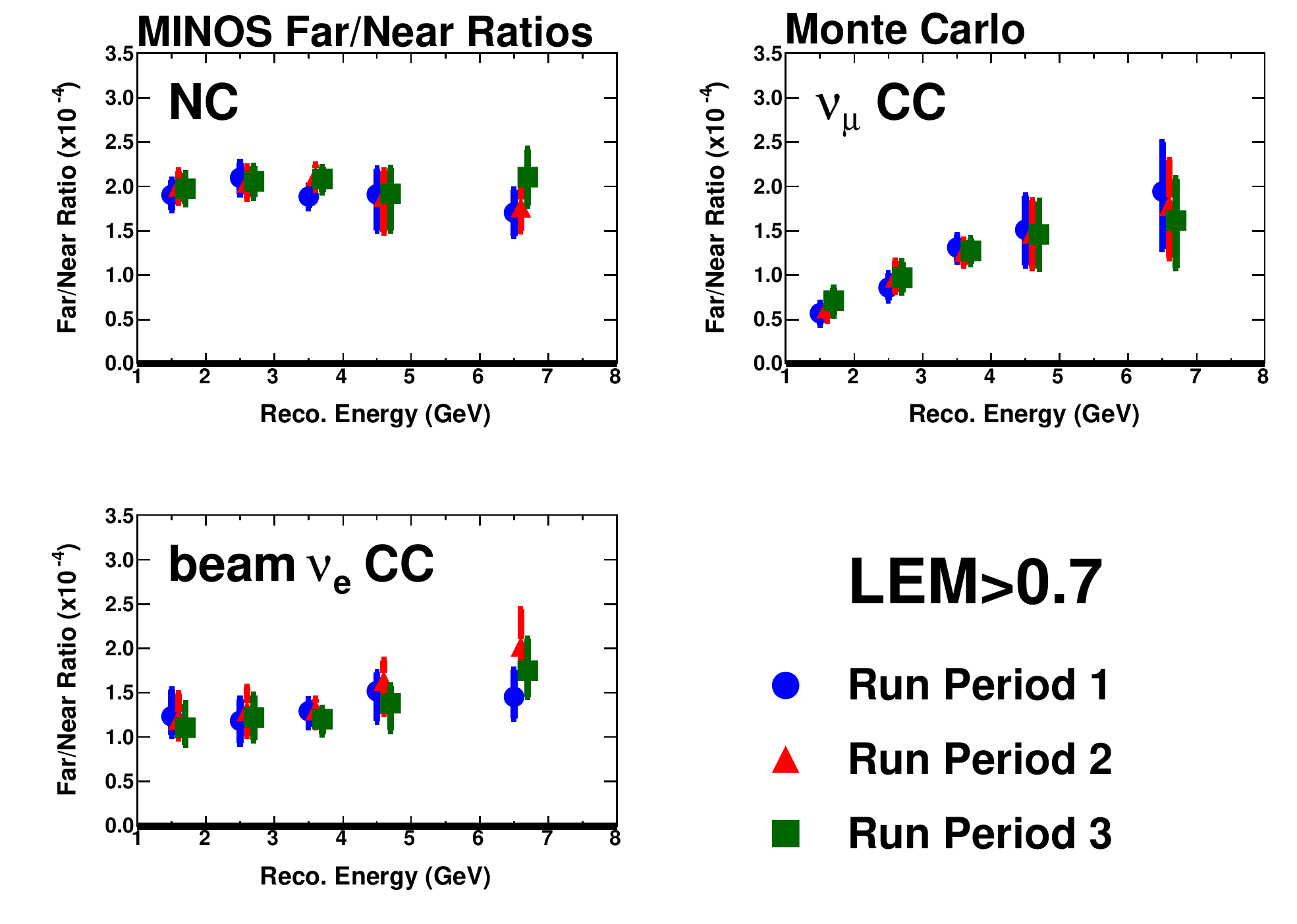}
\caption{The FD-to-ND ratio for the three background event classes as a function of reconstructed energy. These ratios were calculated for events with LEM $>$ 0.7 (the optimal single cut on the discriminant) and are thus for illustration only; the official analysis uses ratios calculated in each bin of the LEM discriminant. The ``Run Periods'' referred to in the plot are periods of experiment running with slightly different conditions; these conditions are accounted for in the simulation, thus explaining why the values of the ratio differ slightly between Run Periods.} \label{FoverN}
\end{figure}

The FD-to-ND ratio in Eq.~\ref{extrapolation} is taken directly from the simulation and is responsible for the extrapolation from ND to FD. It corrects for a variety  of effects, including flux differences ($1/R^2$ fall-off, acceptance, decay kinematics, focusing), fiducial volume effects, energy smearing, oscillations, and detector effects. Similarities between the detectors, including associated similarities between their simulations, allow much of the systematic uncertainty to cancel to first order in this ratio. The resulting values of this ratio, with systematic error bars, are shown in Fig.~\ref{FoverN} for each of the three ND-based background component types, as a function of reconstructed energy.

With this information, we are now able to make the background and signal predictions. In particular, we have the ability to adjust the predictions for different sets of oscillation parameters. To illustrate, Table~\ref{predictions} lists the predicted counts of events with LEM $>$ 0.7 for two reference values of $\theta_{13}$, assuming a commonly used set of values for other oscillation parameters.

\begin{table}[ht]
\begin{center}
\caption{Predicted event counts for events with LEM $>$ 0.7. These numbers were calculated assuming $\delta$ = 0, $\theta_{23} = \pi/4$, and $\Delta m_{32}^2 = 2.32\times10^{-3} \mathrm{eV}^2$ (i.e., normal mass hierarchy). The line labeled ``$\nu_e$ CC'' refers to the intrinsic beam $\nu_e$ component. The nonzero signal in the $\theta_{13}$ = 0 case is due to $\Delta m_{21}^2$-driven oscillations.}
\begin{tabular}{|c|c|c|}
\hline
\multirow{2}{*}{Event Class} & \multicolumn{2}{|c|}{$\sin^2\left(2\theta_{13}\right)$} \\ \cline{2-3}
&  \, \, \, 0 \, \, \, & \, \, \, 0.1\, \, \,\\
\hhline{|=|=|=|}
NC & 34.1 & 34.1\\ \hline
$\nu_\mu$ CC & 6.7 & 6.7\\ \hline
$\nu_e$ CC & 6.4 & 6.2\\ \hline
$\nu_\tau$ CC & 2.2 & 2.1 \\ \hline
\textbf{Total Background} & \textbf{49.4} & \textbf{49.1} \\ \hline
$\boldsymbol{\nu_\mu \rightarrow \nu_e}$ \textbf{CC} & \textbf{0.2} & \textbf{19.1} \\ \hline
\end{tabular}
\label{predictions}
\end{center}
\end{table}

\subsection{The FD Data}

\begin{figure}[ht]
\centering
\includegraphics[width=80mm]{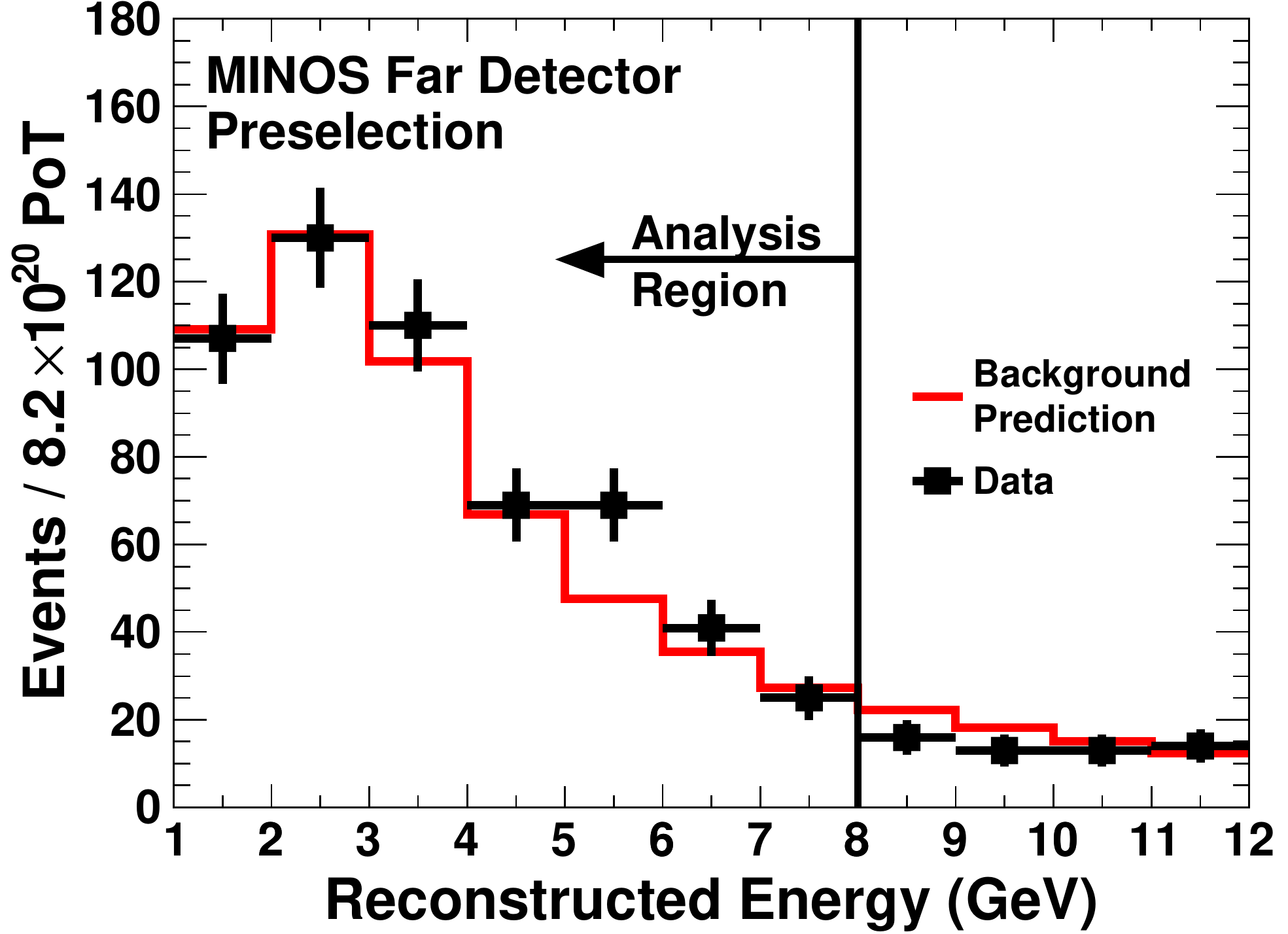}
\caption{Predicted (assuming $\theta_{13}$ = 0) and observed reconstructed energy spectra in the FD.} \label{FDpreselE}
\end{figure}

After applying the fiducial and pre-selection cuts to the events in the FD data, we obtain the energy spectrum shown in Fig.~\ref{FDpreselE}. Although the pre-selection cuts events with reconstructed energy above 8 GeV, we performed the prediction out to 12 GeV and compared it to the data to test the robustness of our background prediction method.

\begin{figure}[ht]
\centering
\includegraphics[width=80mm]{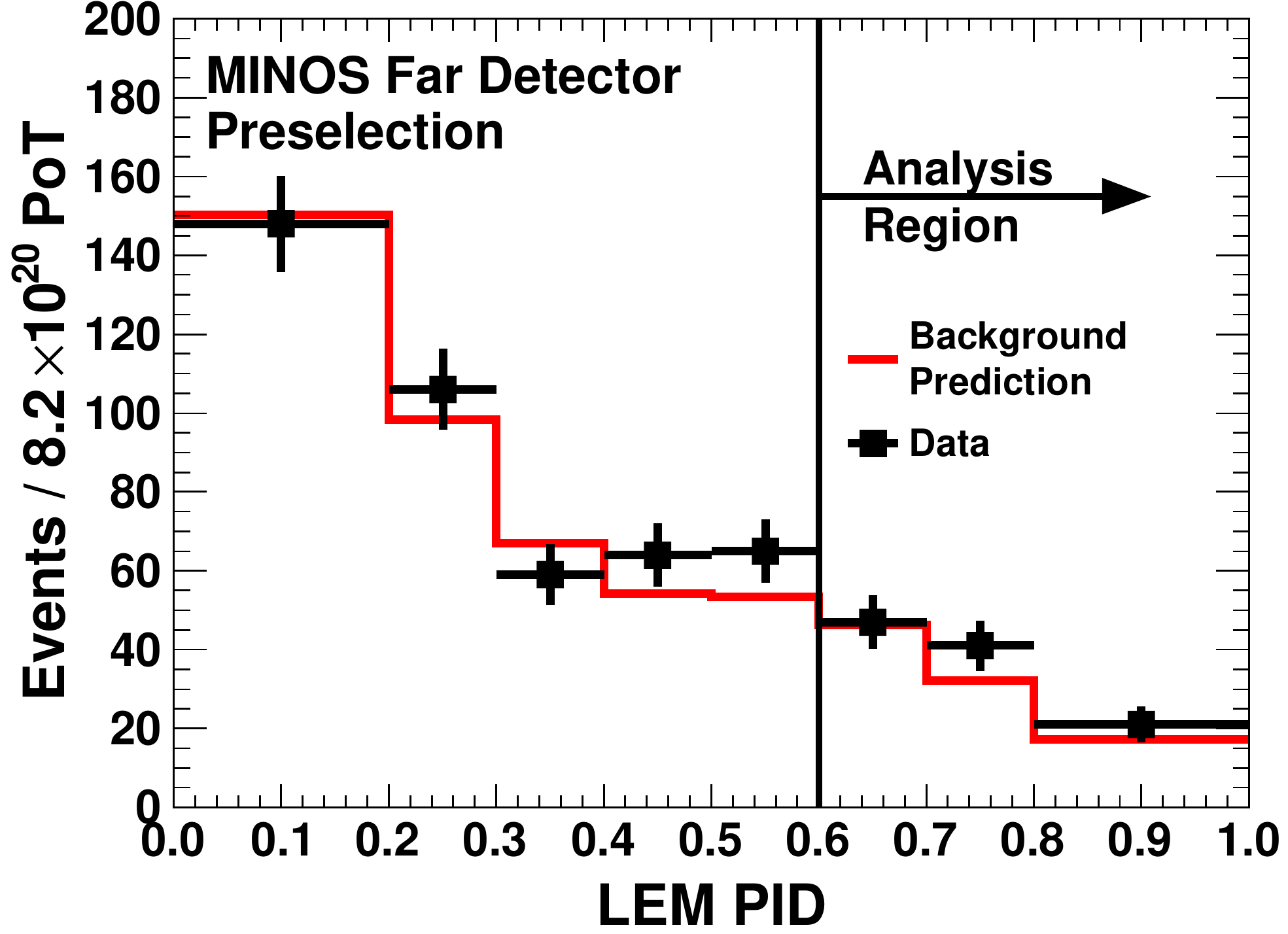}
\caption{Predicted (assuming $\theta_{13}$ = 0) and observed LEM discriminant distributions in the FD.} \label{FDpreselLEM}
\end{figure}

In the LEM discriminant distribution, we expect negligible signal in the LEM $<$ 0.5 region and thus can use it as a test of the entire analysis chain. We predict 370 $\pm$ 19(stat.) background events with LEM $<$ 0.5 and observe 377 -- well within statistical error. Sufficiently assured of the soundness of our analysis method, we proceed to unblind the signal-enhanced region of the LEM discriminant distribution. The full spectrum is shown in Fig.~\ref{FDpreselLEM}. For reference only, we can count the number of events with a LEM discriminant above the optimal cut of 0.7; we predict 49.5 $\pm$ 2.8(syst.) $\pm$ 7.0(stat.) background events and observe 62 events, an excess of approximately 1.7$\sigma$.

\subsection{Fitting to the Data}

\begin{figure*}[ht]
\centering
\includegraphics[width=56mm]{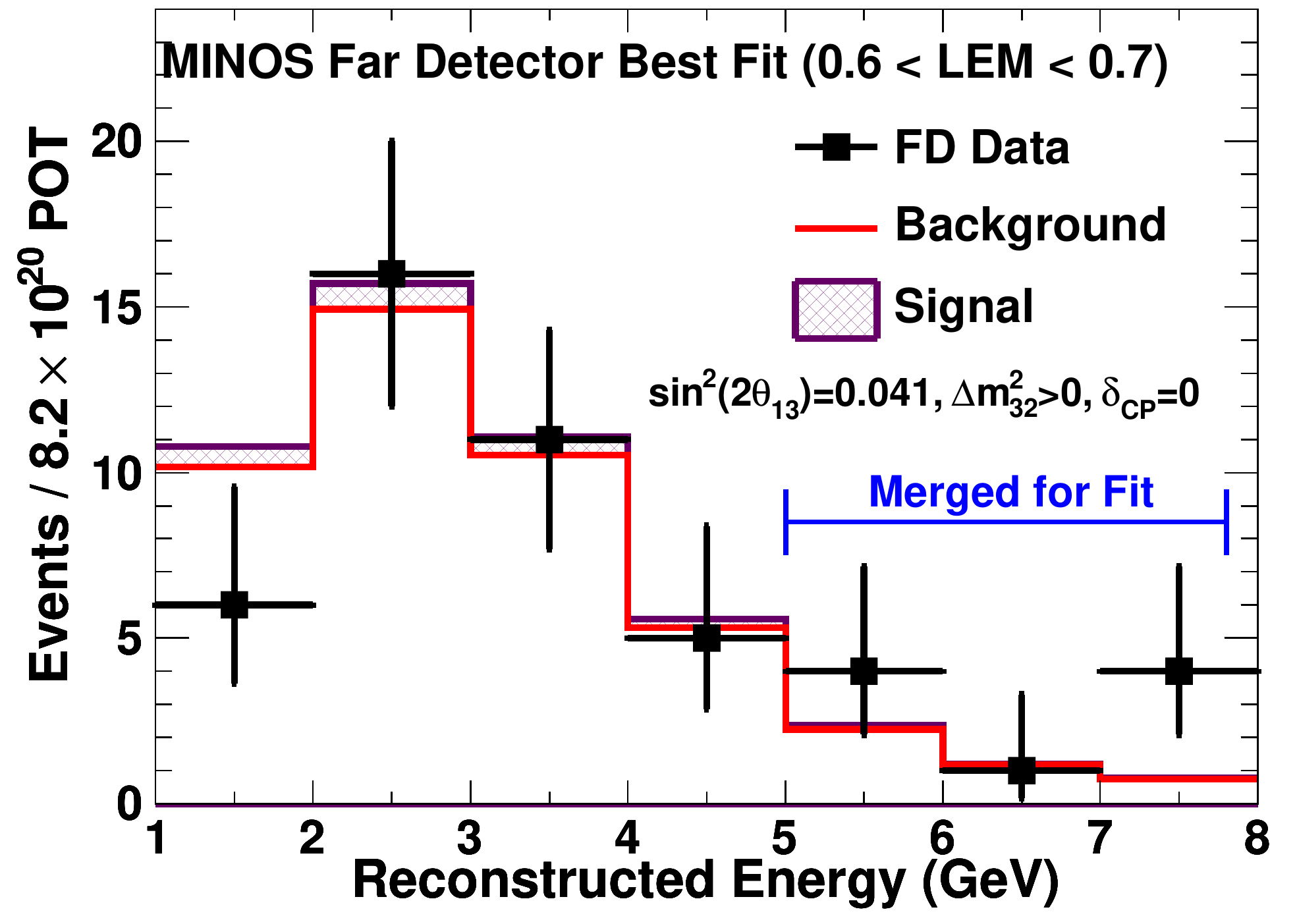}
\includegraphics[width=56mm]{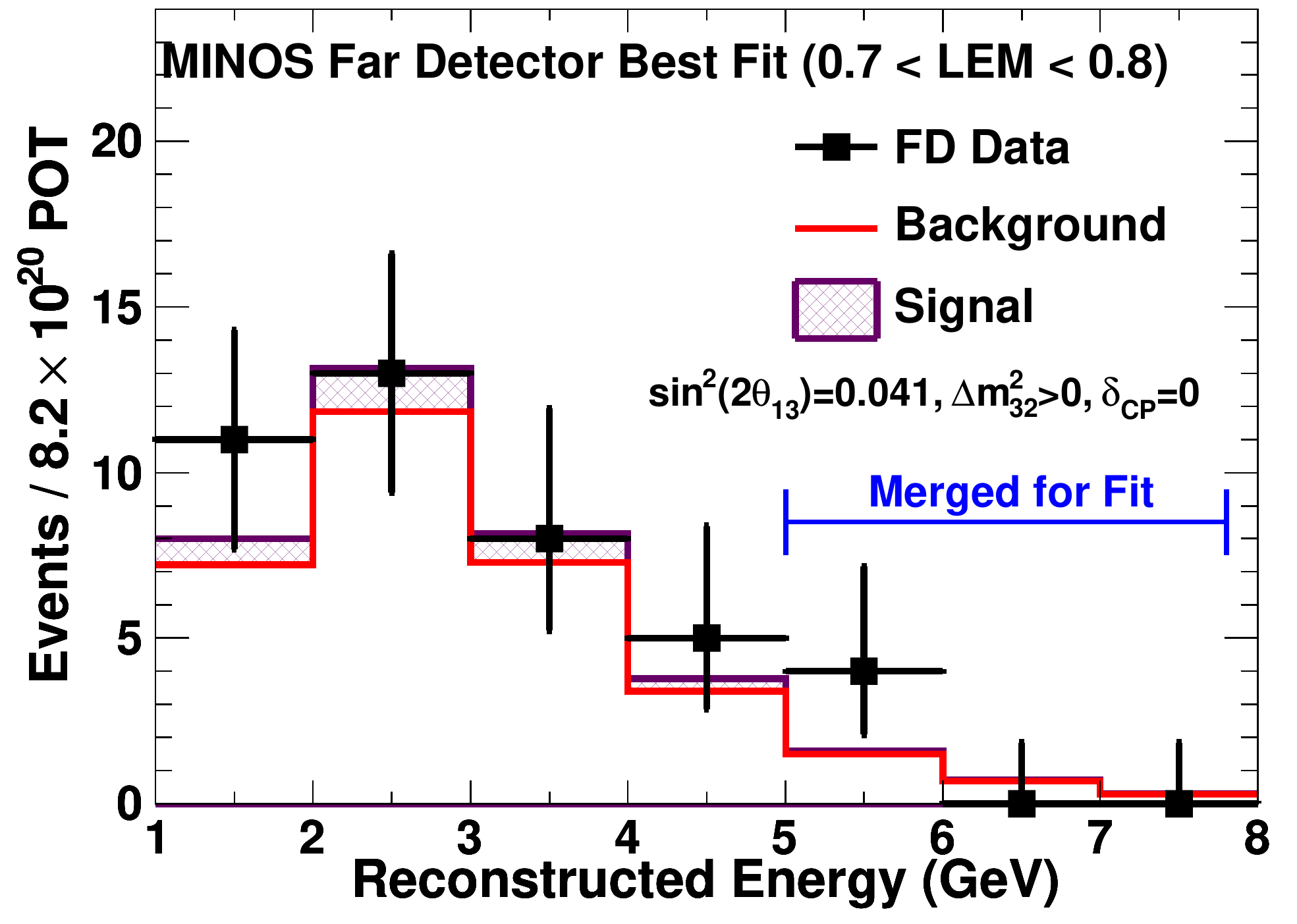}
\includegraphics[width=56mm]{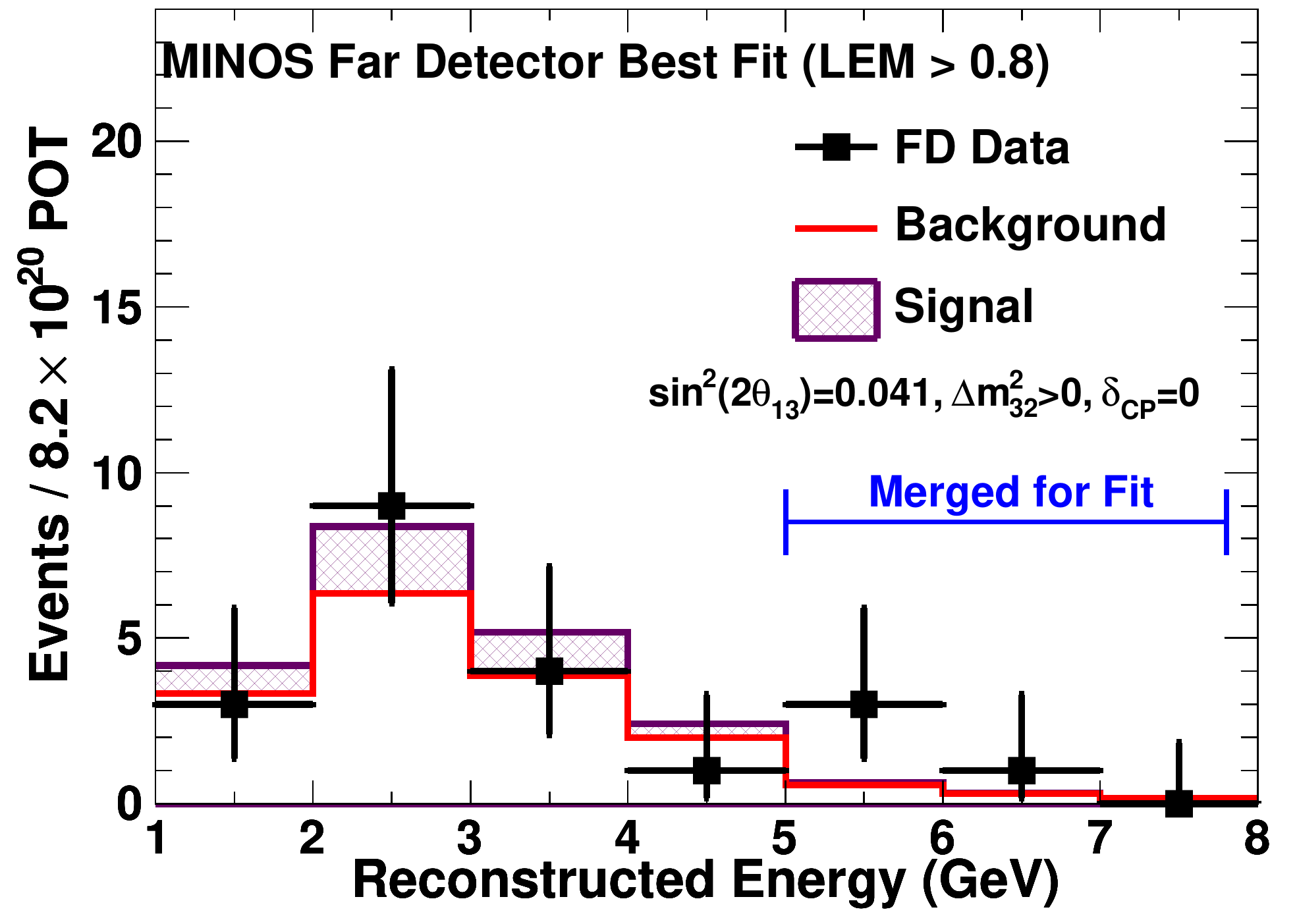}
\caption{Stacked background prediction and best-fit signal assuming $\delta$ = 0, $\theta_{23} = \pi/4$, and $\Delta m_{32}^2 = 2.32\times10^{-3} \mathrm{eV}^2$, with data superimposed. Note that each reconstructed energy spectrum is for a different bin of the LEM discriminant.} \label{bestfit}
\end{figure*}

\begin{figure}[ht]
\centering
\includegraphics[width=80mm]{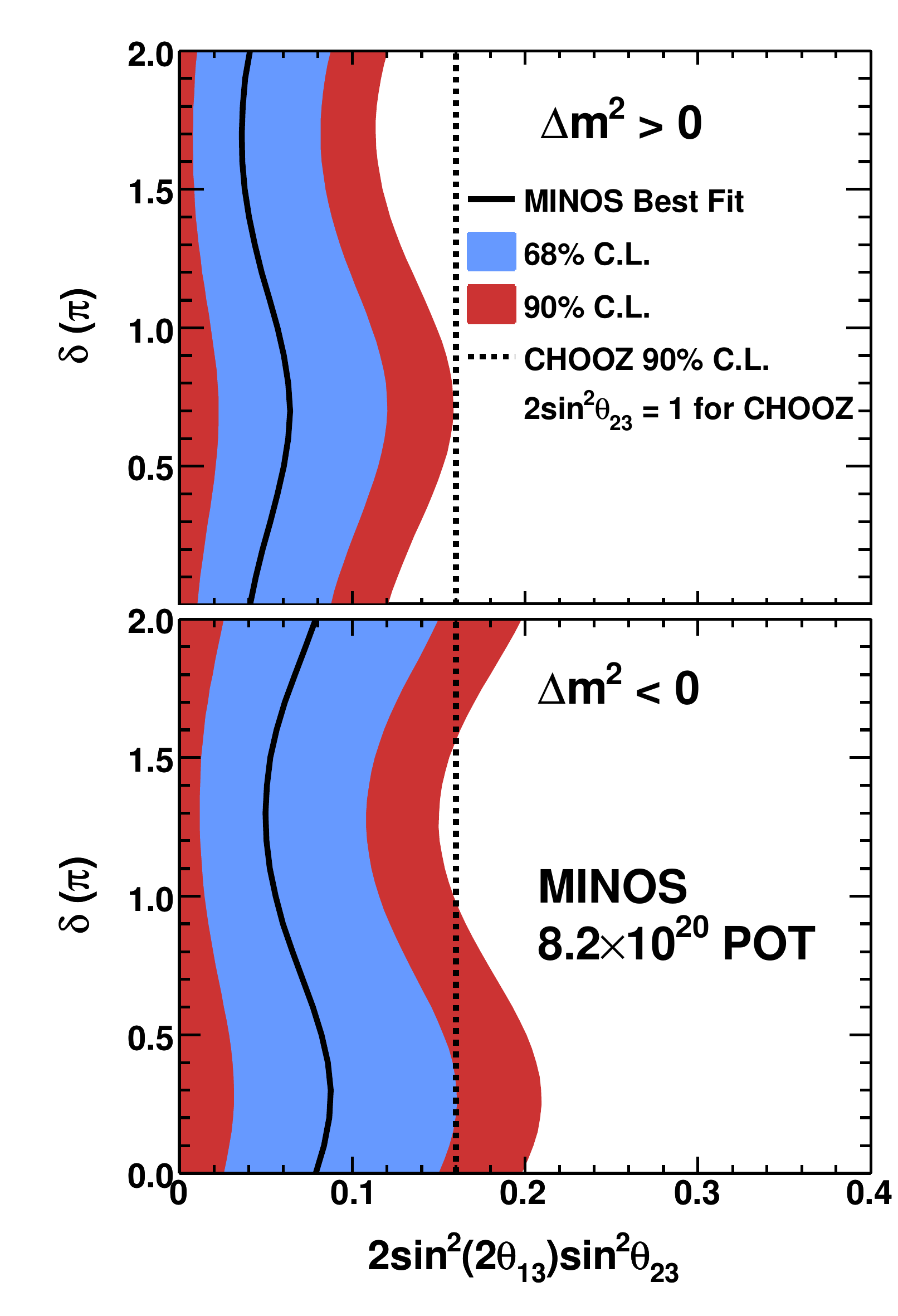}
\caption{Allowed ranges and best fits for $2\sin^2(\theta_{23})\sin^2(2\theta_{13})$ as a function of $\delta$. The upper (lower) panel assumes the normal (inverted) mass hierarchy. The CHOOZ 90\% C.L. upper limit shown assumes $\theta_{23} = \pi/4$ and $\Delta m_{32}^2 = 2.32\times10^{-3} \mathrm{eV}^2$.} \label{contours}
\end{figure}

The official fit to the data is performed in three LEM bins and five reconstructed energy bins. As an example, Fig.~\ref{bestfit} shows the observed reconstructed energy spectra in the three LEM bins superimposed on the best-fit signal and predicted background under the conditions indicated. Our official result, described in Ref.~\cite{MINOSnue3}, is given in Fig.~\ref{contours}. These contours were calculated according to the Feldman-Cousins prescription~\cite{FC} and thus provide correct coverage near the physical boundary at $\theta_{13} = 0$. In addition, uncertainties in the other oscillation parameters have been accounted for in generating these contours. Assuming $\delta = 0$ and $\theta_{23} = \pi/4$, we find that $\sin^2\left(2\theta_{13}\right) = 0.041^{+0.047}_{-0.031}\ \left(0.079^{+0.071}_{-0.053}\right)$ for the normal (inverted) mass hierarchy. Under the same set of assumptions, $\sin^2\left(2\theta_{13}\right) < 0.12 \left(0.20\right)$ at 90\% C.L. for the normal (inverted) mass hierarchy. Finally, these results exclude the null hypothesis of $\sin^2\left(2\theta_{13}\right) = 0$ at 89\% C.L.

\begin{acknowledgments}
We thank DPF and the organizers for the opportunity to present these results at its 2011 meeting. These proceedings report work supported by the U.S.\ DOE; the U.K.\ STFC; the U.S.\ NSF; the State and University of Minnesota; the University of Athens, Greece; and Brazil's FAPESP, CNPq, and CAPES.  We are grateful to the Minnesota Department of Natural Resources, the crew of the Soudan Underground Laboratory, and the staff of Fermilab for their contributions to this effort.
\end{acknowledgments}

\bigskip 

\end{document}